\documentclass[prl,twocolumn,aps,preprintnumbers]{revtex4}
\usepackage{subfigure}
\usepackage{graphicx}
\usepackage{cancel}
\usepackage{amssymb}
\usepackage{textcomp}
\usepackage{amsmath}
\usepackage{bm}
\usepackage{times}
\usepackage{epsfig}
\usepackage{color}
\begin{document}
\preprint{CALT 68-2844}
\title{\Large Higgs Properties and Fourth Generation Leptons}
\author{Koji Ishiwata and Mark B. Wise}
\affiliation{\\  \\
 California Institute of Technology, Pasadena, CA, 91125 USA}
\date{\today}
\begin{abstract}
  It is possible that there are additional vector-like generations
  where the quarks have mass terms that do not originate from weak
  symmetry breaking, but the leptons only get mass through weak
  symmetry breaking. We discuss the impact that the new leptons have
  on Higgs boson decay branching ratios and on the range of allowed
  Higgs masses in such a model (with a single new vector-like
  generation). We find that if the fourth generation leptons are too
  heavy to be produced in Higgs decay, then the new leptons reduce the
  branching ratio for $h \rightarrow \gamma \gamma$ to about 30\% of
  its standard-model value. The dependence of this branching ratio on
  the new charged lepton masses is weak. Furthermore the expected
  Higgs production rate at the LHC is very near its standard-model
  value if the new quarks are much heavier than the weak scale. If the
  new quarks have masses near the cutoff for the theory, then for
  cutoffs greater than $10^{15}~{\rm GeV}$, the new lepton masses
  cannot be much heavier than about $100~{\rm GeV}$ and the Higgs mass
  must have a value around $175~{\rm GeV}$.
\end{abstract}
\maketitle
\section{I. Introduction}

We have observed three generations of quarks and leptons, however,
there is no convincing prediction for the number of generations that
exist. Hence examining the physics of extensions of the standard model
with additional generations of quarks and leptons is worthwhile.
Experimental constraints on fourth generation of quark masses are very
strong, $m_{u',d'} \gtrsim 330~{\rm
  GeV}$~\cite{Aaltonen:2009nr,Lister:2008is}. To be consistent with
these constraints, an extension of the minimal standard model with a
chiral fourth generation of quarks and leptons (and no other degrees
of freedom) must be low energy effective theory with a cutoff not far
from the TeV scale. This is because the large fourth generation quark
Yukawa couplings grow with energy scale and one encounters a Landau
pole after only a modest amount of renormalization group
evolution. Furthermore there are issues with stability of the Higgs
potential in such a model.  Experimental limits on the masses of
fourth generation leptons, on the contrary, are much less stringent.
Heavy charged lepton masses must be larger than about $100~{\rm GeV}$,
while stable (unstable) heavy neutral leptons must be heavier than
about $45~{\rm GeV}$ ($90~{\rm GeV}$)~\cite{Nakamura:2010zzi}.

In this paper an additional vector-like fourth generation ({\it i.e.},
a chiral fourth generation plus its mirror) is considered.  In this
framework, one can construct scenarios where the fourth generation
quarks get a mass term that does not require weak symmetry breaking,
but the leptons are forbidden from getting such a mass term.  The
model constructed in Ref.~\cite{Perez:2011pt} where baryon and lepton
number are gauged and spontaneously broken is an example of this. In
such models over most of the parameter space fourth generation quarks
acquire masses much greater than the weak scale, nonetheless, their
Yukawa couplings to the Higgs doublet can be small.  On the other
hand, fourth generation leptons cannot have masses far above the weak
scale and fourth generation lepton masses around $100~{\rm GeV}$ are
reasonable. Hence, the problems associated with Landau poles not far
from the weak scale and vacuum stability do not occur over a wide
range of the allowed parameter space.

Even if the fourth generation quarks are very heavy, the new leptons
have a dramatic effect on the decays of the Higgs boson. (For a study
of Higgs physics in four generation models, see
Ref.~\cite{Kribs:2007nz}.)  Since the fourth generation quarks have
mass terms that do not require weak symmetry breaking, they decouple
as their masses increase.  If the new quarks are much heavier than the
weak scale then the Higgs production rate at the LHC is near its
standard-model value, but we find that the $h \rightarrow \gamma
\gamma$ branching ratio is reduced to about 30\% of its standard-model
value. This reduction depends weakly on the charged lepton masses and
so it is a signature for this scenario.

Although the small fourth generation quark Yukawa couplings do not
develop Landau poles below the GUT or Planck scale, the new leptons
may give rise to Landau poles in coupling constants or a vacuum
instability in Higgs potential. We study the impact that these leptons
have on the renormalization group evolution of the Yukawa couplings
and the Higgs self-coupling.  The Higgs mass squared is proportional
to its self-coupling $\lambda$.  There are upper and lower bounds on
the Higgs mass from the requirement that $\lambda(\mu) <\infty$ (no
Landau pole) and $\lambda(\mu)>0$ (vacuum stability) for scales $\mu$
less than the cutoff of the theory. We find the Higgs mass (denoted as
$m_h$) must be around $175~{\rm GeV}$ and the fourth generation lepton
masses cannot be greater than about $100~{\rm GeV}$ when the cutoff
$\Lambda_c\simeq10^{15}~{\rm GeV}$, assuming the fourth generation
quark masses are at the cutoff.  With a low cutoff of $10~{\rm TeV}$,
the Higgs mass must be in the range $120~{\rm GeV} \lesssim m_h
\lesssim 400~{\rm GeV}$ and roughly speaking the fourth generation
lepton masses should be smaller than the Higgs mass.

While we were working on this paper, Ref.~\cite{Knochel:2011ng}
appeared. The research presented here is similar to that in 
Ref.~\cite{Knochel:2011ng}, however this paper is focused on a
particular class of models

\section{II. The model}
\label{sec:model}

We consider a model with a vector-like fourth generation of quarks and
leptons in addition to the standard-model particles.  This fourth
generation has the $SU(2)$ left-handed quark doublet, $Q_L'=(u_L',
d_L')$, right-handed up- and down-type quark singlets, $u_R'$ and
$d_R'$, left-handed lepton doublet, $L_L'=(\nu_L', e_L')$, and
right-handed charged and neutral lepton singlets, $e_R'$ and
$\nu_R'$. The mirror fourth generation particles are: the $SU(2)$
right-handed quark doublet, $Q_R''=(u_R'', d_R'')$, left-handed up-
and down-type quark singlets, $u_L''$ and $d_L''$, the right-handed
lepton doublet, $L_R''=(\nu_R'', e_R'')$, and the left-handed charged
and neutral leptons singlets, $e_L''$ and $\nu_L''$.  Their $U(1)_Y$
charges are the same as the existing fermions in the standard model.

In addition to their gauge invariant kinetic terms, the new quarks
have the following mass terms and Yukawa couplings to the Higgs
doublet denoted as $H$,
\begin{eqnarray}
\label{quark}
&&\Delta {\cal L} _{\rm q}
=-M_Q {\bar Q'_L} Q''_R-M_U{\bar u}_R' u_L'' -M_D{\bar d}_R' d_L'' \nonumber \\
&&-h_U' {\bar Q'_L} \epsilon H^*u_R'-h_U''{\bar Q}_R'' \epsilon H^* u_L''
 \nonumber \\
&&-  h_D' {\bar Q'_L}  H d_R'-h_D'' {\bar Q''_R}  H d_L''+{\rm h.c.},
\label{Lquark}
\end{eqnarray}
where $\epsilon$ is the antisymmetric $2\times2$ matrix (in weak
$SU(2)$ space) with non zero components, $\epsilon_{12}=1$ and
$\epsilon_{21}=-1$. Here we ignore terms which mix the fourth
generation quarks with the familiar quarks for simplicity.  However,
we imagine that there are small couplings of this type that allow the
new quarks to decay.  (For constraints on the mixings in quark sector,
as well as lepton sector, from experiments, see {\it e.g.},
Ref.~\cite{Kribs:2007nz}.) Such mixings are forbidden if there is a
global $U(1)$ symmetry where the fourth generation quarks and the
ordinary quarks have different charge.  If the breaking of this
symmetry is small, then the mixings are expected to be small.

For the new leptons, the analogous couplings are given by
\begin{eqnarray}
\label{lepton}
&&\Delta{\cal L}_{\rm l}
=-h_E' {\bar L_L'} H e_R'-h_E'' {\bar L_R''} He_L''\nonumber \\
&&-h_N {\bar L_L'}\epsilon H^* \nu_R'-h_N'' {\bar L_R''}\epsilon H^* \nu_L'' 
+{\rm h.c.}.
\end{eqnarray}
The crucial difference between the quark and lepton sectors is the
absence of bare lepton mass terms. The bare mass terms, in addition to
the terms which mix the new leptons and ordinary leptons, can be
forbidden if one assumes a global $U(1)$ symmetry where the primed
leptons, double primed leptons and ordinary leptons have different
charge. When the charge of the ordinary leptons under this symmetry is
zero, then the ordinary neutrinos can have Majorana masses.  If there
is no mixing between the fourth generation and ordinary leptons, one
can have an acceptable scenario with stable fourth generation
neutrinos.

Usually one does not impose global symmetries since quantum gravity
effects will violate them. However, it is possible that there are
underlying gauge symmetries that leave Eqs.~(\ref{quark}) and
(\ref{lepton}) as the low energy effective theory after they
spontaneously break.  In fact baryon and lepton number could be such
gauge symmetries. By adding a vector-like fourth generation one can
gauge baryon and lepton number provided the difference in baryon
number between the fourth generation and mirror generation is $-1$ and
the difference in lepton number between the fourth generation and
mirror generation is $-3$~\cite{Perez:2011pt}. The quarks in these
families have mass terms that do not require weak symmetry breaking
provided we introduce a scalar, $S_B$, with baryon number $1$ that
gets a vacuum expectation value (VEV).  Similarly fourth generation
lepton masses that do not require weak symmetry breaking arise if a
scalar, $S_L$, with lepton number $3$ gets a VEV. However this charge
for $S_L$ is not preferred since then one cannot generate light
neutrino masses through the seesaw mechanism~\cite{TypeI}. If we use
$S_L$ with lepton number $2$ to break lepton number, Majorana neutrino
masses for the light neutrinos are generated through the seesaw
mechanism and furthermore proton decay is forbidden since the field
that breaks lepton number has even charge.  In that case the breaking
of lepton number does not give rise to mass terms for the fourth
generation leptons.

It is appropriate to keep the fourth generation quarks in the low
energy effective theory if their masses are well below the scale of
baryon number symmetry breaking. However, there is no particular
reason for this to be the case and the generic situation is that one
would only be left with just the fourth generation leptons and the
standard-model particles in the low energy effective theory.

\section{III. Phenomenology of the Higgs boson}

Here we discuss phenomenology of the Higgs boson in the model with the
vector-like fourth generation. In this model the fourth generation
quark and lepton masses have different origins.  Thus they have
different impact on the properties of the Higgs boson. In this section
we discuss Higgs boson production and decay at the LHC.

\subsection{A. Fourth generation quarks and the Higgs production}
  
Here we study the impact of vector-like fourth generation on Higgs
production at the LHC. In the usual chiral fourth generation scenario,
the Higgs production rate, which is dominated by the gluon fusion
process, is increased by about a factor of nine and this result is
almost independent of the chiral fourth generation quark masses. (See
Ref.~\cite{Kribs:2007nz} and references therein.)  Contrary to such a
result, we will see that the production rate rapidly approaches to the
standard-model value as the vector-like fourth generation quark masses
get larger than a TeV.

Let us derive the interaction Lagrangian of fourth generation mass
eigenstate quarks with the Higgs boson. It is convenient to introduce
the four-component fourth generation quark fields: $\psi_U'$,
$\psi_U''$, $\psi_D'$ and $\psi_D''$. Their left and right components
are $\psi_{UL}'=u_L'$, $\psi_{UR}'=u_R''$, $\psi_{UL}''=u_L''$,
$\psi_{UR}''=u_R'$ and similarly for the down-type quarks.  In the
basis $\Psi_U= (\psi'_U,\psi''_U)$ (and including the effects of weak
symmetry breaking) the up-type quark mass terms are from
Eq.~(\ref{Lquark}) as,
\begin{equation}
\Delta {\cal L}^{(u)}_{\rm mass}
=-{\bar \Psi_{UL}}{\cal M}_U \Psi_{UR} +{\rm h.c.},
\end{equation}
where
\begin{equation}
{\cal M}_U
=\left(\begin{array}{cc}M_Q & m'_U \\ m''_U & M_U\end{array}\right),
\end{equation}
and $m'_U=h_U' v/\sqrt{2}$ and $m''_U=h_U''^*v/\sqrt{2}$. Here
$\langle H^0 \rangle =v/\sqrt{2}$ with $v\simeq 246~{\rm GeV}$, the VEV
of the neutral component of the Higgs doublet.  The up-type quark mass
matrix is diagonalized by making unitary transformations $V_L(u)$ and
$V_R(u)$ on the left and right-handed up-type quark fields so that,
\begin{equation}
V_L(u)^{\dagger}{\cal M}_UV_R(u)
=\left(\begin{array}{cc}M_{U_1} & 0 \\ 0 & M_{U_2}\end{array}\right).
\end{equation}
We denote the two up-type quark mass eigenstates as $U_1$ and $U_2$
and take $M_{U_1}>M_{U_2}$. The mass eigenvalues are,
\begin{equation}
M_{U_{1,2}}^{2}=
{1 \over 2} \left[ ( M_Q^2+m'^{2}_U   +M_U^2 +m''^{2}_U)
 \pm  \sqrt{X+Y} \right],
\end{equation}
with
\begin{eqnarray}
X&=&\left(M_Q^2+m_U'^{2} - M_U^2 -m_U''^{2}\right)^2, \\
Y&=&4\left( m_U'' M_Q+ m_U'M_U  \right)^2.
\end{eqnarray}
For simplicity we assume the up-type quark Yukawas, $h_U^{\prime}$ and
$h_U^{\prime \prime}$, are real so that the transformations that
diagonalize the mass matrix are the real orthogonal matrices,
\begin{eqnarray}
V_L(u)&=&\left(
\begin{array}{cc}{ \rm cos}~\theta^{(u)}_L &{ \rm sin}~\theta^{(u)}_L \\ 
-{ \rm sin}~\theta^{(u)}_L & { \rm cos}~\theta^{(u)}_L
\end{array}\right),
\\
V_R(u)&=&\left(
\begin{array}{cc}{ \rm cos}~\theta^{(u)}_R &{ \rm sin}~\theta^{(u)}_R \\ 
-{ \rm sin}~\theta^{(u)}_R & { \rm cos}~\theta^{(u)}_R
\end{array}\right).
\end{eqnarray}
Since $M_Q$ and $ M_ U$ are larger than the mass terms that arise from
weak symmetry breaking, the angles $\theta^{(u)}_{L,R}$ are small and
their cosines are positive. The angles are given by,
\begin{equation}
{\rm tan}~\theta_L^{(u)} 
={m_U''M_Q+m_U'M_U \over M_{U_2}^{2}   -M_Q^2-m_U'^{2}  },
\label{eq:thetaL}
\end{equation}
and the right-handed angle is given by flipping $m'$ with $m''$,
\begin{equation}
{\rm tan}~\theta_R^{(u)} 
={m_U'M_Q+m_U''M_U \over M_{U_2}^{2}   -M_Q^2-m_U''^2  }.
\label{eq:thetaR}
\end{equation}
Similar formulae hold for the down-type fourth generation quarks.

For the calculation of the Higgs production rate, we need to know the
couplings of the Higgs boson to the fourth generation quark mass
eigenstates which do not change the type of heavy quark.  Using the
above definitions, these are
\begin{eqnarray}
{\cal L}^{Q}_{\rm Higgs}
&=&- {\mu_{U_1} \over v} h{\bar U}_1 U_1 
- {\mu_{U_2} \over v} h{\bar U}_2 U_2 
\nonumber \\
&-&
{\mu_{D_1} \over v} h{\bar D}_1 D_1 
- {\mu_{D_2} \over v} h{\bar D}_2 D_2 ,
\end{eqnarray}
where
\begin{eqnarray} 
\mu_{U_1}&=&-{{\rm cos}~\theta_L^{(u)}~{\rm cos}~\theta_R^{(u)}}
\left( m_U'{\rm tan}~\theta_R^{(u)}+m_U''{\rm tan}~\theta_L^{(u)}\right),
\nonumber \\ 
\mu_{U_2}&=&{{\rm cos}~\theta_L^{(u)}~{\rm cos}~\theta_R^{(u)} }
\left( m_U'{\rm tan}~\theta_L^{(u)}+m_U''{\rm tan}~\theta_R^{(u)}\right),
\nonumber \\
\mu_{D_1}&=&-{{\rm cos}~\theta_L^{(d)}~{\rm cos}~\theta_R^{(d)}}
\left( m_D'{\rm tan}~\theta_R^{(d)}+m_D''{\rm tan}~\theta_L^{(d)}\right),
\nonumber \\
\mu_{D_2}&=&{{\rm cos}~\theta_L^{(d)}~{\rm cos}~\theta_R^{(d)}}
\left( m_D'{\rm tan}~\theta_L^{(d)}+m_D''{\rm tan}~\theta_R^{(d)}\right).
\nonumber \\ 
\label{eq:muD-}
\end{eqnarray}
For comparison we give the coupling of the Higgs boson to the top quark,
\begin{equation}
{\cal L}^t_{\rm Higgs}=-{m_t \over v} h {\bar t } t,
\end{equation}
where $m_t$ is the top quark mass.

Now we are ready to calculate the Higgs production rate at the LHC. As
we mentioned, the production rate is dominated by the gluon fusion
process. It is induced by a quark loop. In the standard model a top
quark loop is the main contribution.  In our model, fourth generation
quarks are additional contributions.  Thus the Higgs production rate
in our model divided by its standard-model value is given by
\begin{eqnarray}
  &&{\sigma(gg \rightarrow h) \over \sigma^{\rm SM}(gg \rightarrow h)}
  \nonumber \\
  &&=\left| 1+\sum_{i=1,2} \Bigg[ { \mu_{U_i}\over M_{U_i}} 
    {I \left(r_{U_i}\right)}
   + { \mu_{D_i}\over M_{D_i}}{I \left(r_{D_i}\right) } \Biggr]/ I(r_t)
  \right|^2,
  \nonumber \\
\label{eq:hgg}
\end{eqnarray}
where $r_t\equiv m_h^2/4 m_t^2$, $r_{U_i,D_i}\equiv m_h^2/4
M_{U_i,D_i}^2$ and the function $I(x)$ is
\begin{equation}
I(x)=2[x+(x-1)f(x)]/x^2,
\end{equation}
with
\begin{eqnarray}
f(x)= \left\{
\begin{array}{lc}
{\rm Arcsin}^2(\sqrt{x}) & 0 <x \le 1 \\
-\frac{1}{4}\left[ \log \frac{1+\sqrt{1-1/x}}{1-\sqrt{1-1/x}}-i \pi\right]^2
& 1<x 
\end{array} \right. .
\end{eqnarray}
We neglect the other light quarks, $b,c,s,d$ and $u$.

\begin{figure}[t]
  \begin{center}
    \includegraphics[scale=1.4]{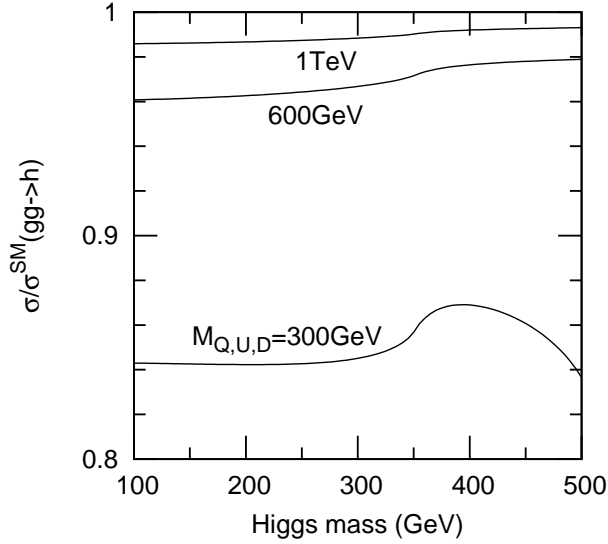}
  \end{center}
  \caption{Ratio of the cross section for $gg\rightarrow h$ in our
    model to its standard-model value as a function of the Higgs
    mass. Here we take $m_U'=2m_U''=60~{\rm GeV}$,
    $m_D'=2m_D''=60~{\rm GeV}$, and plot for $M_Q =M_U=M_D=300~{\rm
      GeV}$, $600~{\rm GeV}$ and $1~{\rm TeV}$ from bottom to top. }
  \label{fig:hgg}
\end{figure}

In Fig.~\ref{fig:hgg} we plot this ratio of the cross sections for
$m_U'=2m_U''=60~{\rm GeV}$, $m_D'=2m_D''=60~{\rm GeV}$ and $M_Q
=M_U=M_D=300~{\rm GeV}$, $600~{\rm GeV}$ and $1~{\rm TeV}$ as a
function of $m_h$.  The production rate rapidly approaches the
standard-model rate as $M_{U,D,Q}$ increase.  This is because the
contribution of the fourth generation quarks is suppressed by
$m'^2_{U,D}/M^2_{Q,U,D}$, $m''^2_{U,D}/M^2_{Q,U,D}$, which can be seen
from Eqs.~(\ref{eq:thetaL})-(\ref{eq:muD-}) and (\ref{eq:hgg}). In
usual fourth generation scenario ({\it i.e.}, not vector-like
scenario), on the contrary, the Higgs production rate by gluon fusion
process is increased by about a factor of nine over the standard-model
production rate and its dependence on the chiral fourth
  generation quark masses is weak.  Therefore, the exclusion of the
Higgs with mass in the range, $131~{\rm GeV} \le m_h \le 204~{\rm
  GeV}$, by the Tevatron in the usual fourth generation
scenario~\cite{Aaltonen:2010sv}, is not applicable to our model.

\subsection{B. Fourth generation leptons and the Higgs decay}

So far we have discussed the impact of a fourth vector-like generation
on the Higgs production rate at the LHC. Next, we study the Higgs
decay branching ratios. The partial decay widths of each mode in our
model are the same as those in standard model, except for the modes
$h\rightarrow gg$, $\gamma Z$ and $\gamma\gamma$ where the decay is
induced by gauge-boson and heavy-fermion loops (and of course if the
new leptons are light enough there are now also Higgs decays to
particles not in the standard model).  We will see shortly that the
decay rate for $h\rightarrow \gamma\gamma$ is significantly changed by
the fourth generation leptons.

The new leptons only get mass through their coupling to the Higgs
doublet, as seen in Eq.~(\ref{lepton}). Then the mass terms in the
new lepton sector are given by,
\begin{eqnarray}
{\cal L}_{\rm mass}^{(l)}=
 -m_E' \bar{e}'e' -m_E'' \bar{e}''e''
 -m_N' \bar{\nu}'\nu' -m_N'' \bar{\nu}''\nu'',
\end{eqnarray}
where $m_E'=h_E'v/\sqrt{2}$, $m_E''=h_E''v/\sqrt{2}$,
$m_N'=h_N'v/\sqrt{2}$ and $m_N''=h_N''v/\sqrt{2}$.  Here $e'$, $e''$,
$\nu'$ and $\nu''$ are all Dirac fermions. Their couplings with the
Higgs boson, therefore, are given in a form similar to the
standard-model fermions:
\begin{eqnarray}
{\cal L}_{\rm Higgs}^{L}&=&
-\frac{m'_E}{v}h\bar{e}'e'-\frac{m''_E}{v}h\bar{e}''e''
\nonumber \\
&-& \frac{m'_N}{v}h\bar{\nu}'\nu'-\frac{m''_N}{v}h\bar{\nu}''\nu''.
\end{eqnarray}
With this Lagrangian, the partial decay width for $h\rightarrow \gamma
\gamma$ in our model is then easily  calculated to be,
\begin{equation}
\Gamma (h \rightarrow \gamma \gamma)=
\frac{\alpha^2G_F m^3_h}{128\sqrt{2}\pi^3} |J_{\gamma\gamma}|^2,
\end{equation}
where $\alpha$ and $G_F$ are fine structure constant Fermi constant
and
\begin{eqnarray}
J_{\gamma\gamma}&=& \left(\frac{2}{3}\right)^2 N_c  I(r_t)
+  I(r_{E'}) +I(r_{E''}) +K(r_W)
\nonumber \\
&+&  \left(\frac{2}{3}\right)^2 N_c
\sum_{i=1,2} { \mu_{U_i}\over M_{U_i}} 
I \left(r_{U_i}\right)
\nonumber \\
&+& \left(-\frac{1}{3}\right)^2 N_c \sum_{i=1,2}  
{ \mu_{D_i}\over M_{D_i}} 
I \left(r_{D_i}\right).
\label{eq:AMPhgamgam}
\end{eqnarray}
Here we explicitly wrote color factor $N_c=3$ and used $r_{E'}\equiv
m_h^2/4 m'^2_{E}$, $r_{E''}\equiv m_h^2/4 m''^2_{E}$ and $r_W \equiv
m_h^2/4m^2_W$ ($m_W$ is $W$ boson mass). The function $K(x)$ is given
by
\begin{eqnarray}
K(x)=-\left[2+3/x+3(2x-1)/x^2 f(x)\right].
\end{eqnarray}
The terms in the first line of Eq.~(\ref{eq:AMPhgamgam}) are coming
from the top quark, two fourth generation charged leptons, and the $W$
boson loop. The rest are from fourth generation quarks.  In the
amplitude, the fourth generation lepton gives a comparable
contribution to the top quark, while the effect of the fourth
generation quarks are very small if $M_{U,D,Q}> 1~{\rm TeV}$.

\begin{figure}[t]
  \begin{center}
    \includegraphics[scale=1.4]{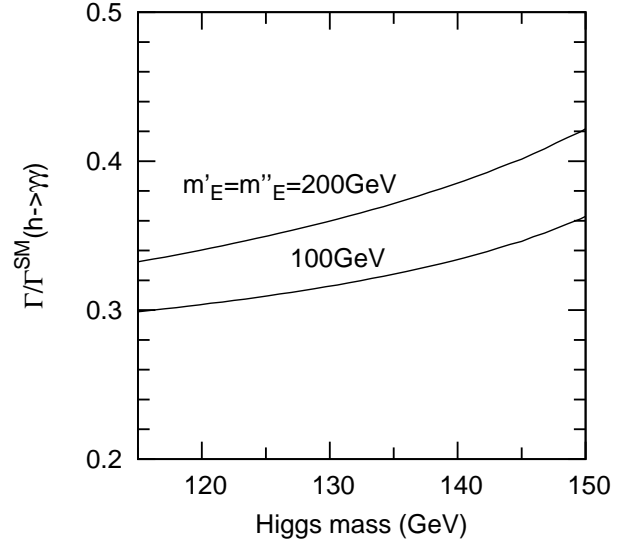}
  \end{center}
  \caption{Ratio of $\Gamma(h\rightarrow \gamma\gamma)$ to its
    standard-model value. Here we take $m_E'=m_E''=100~{\rm GeV}$ and
    $200~{\rm GeV}$ from bottom to top.}
  \label{fig:hgamgam}
\end{figure}

\begin{figure}[t]
  \begin{center}
    \includegraphics[scale=0.75]{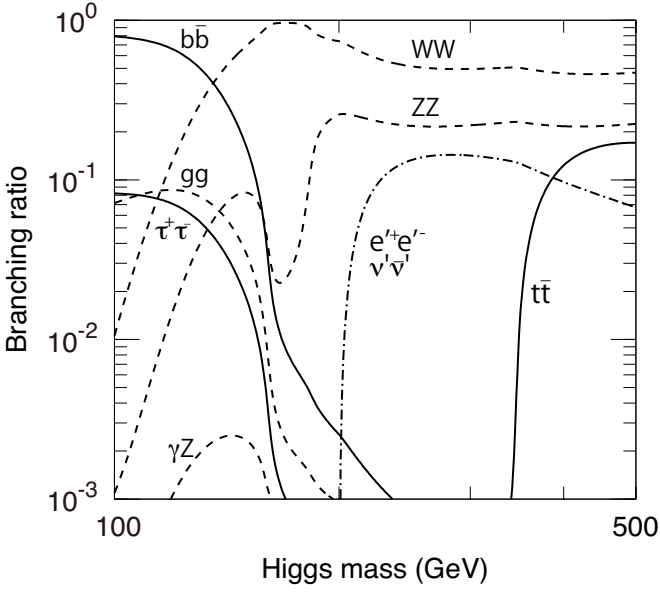}
  \end{center}
  \caption{Branching ratios of the Higgs decay. Here we take
    $m_E'=m_E''=100~{\rm GeV}$ and $m_{N}'=m_{N}''=100~{\rm GeV}$. The
    modes in which the final state is standard-model fermion pair are
    drawn in solid line ($t\bar{t}$, $b{\bar b}$ and $\tau^+ \tau^-$),
    and the modes in which the final state is standard-model gauge
    bosons, {\it i.e.}, $W^+W^-$, $ZZ$, $\gamma\gamma$, $\gamma Z$ and
    $gg$, are in dashed lines. Here we omit $c \bar{c}$ line. The
    branching ratios of new leptons, $e'^+e'^-$, $e''^+e''^-$,
    ${\nu}'\bar{\nu}'$ and ${\nu}''\bar{\nu}''$, are drawn in
    dot-dashed lines. In this plot, they all are identical. The line
    shows the branching ratio for $e'^+e'^-$ plus $e''^+e''^-$
    (${\nu}'\bar{\nu}'$ plus ${\nu}''\bar{\nu}''$) with sum denoted
    just by ``$e'^+e'^-$'' (``${\nu}'\bar{\nu}'$'').}
  \label{fig:sm4i}
\end{figure}

\begin{figure}[t]
  \begin{center}
    \includegraphics[scale=0.75]{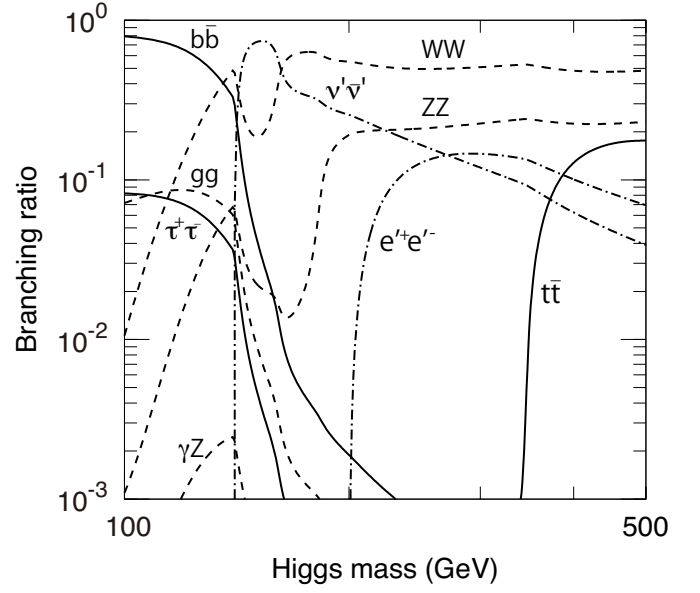}
  \end{center}
  \caption{The same as Fig.~\ref{fig:sm4i}, except for taking
    $m_N'=m_N''=70~{\rm GeV}$. The branching ratios of $e'^+e'^-$ and
    $e''^+e''^-$ ($\nu'\bar{\nu}'$ and $\nu''\bar{\nu}''$) are
    equal. }
  \label{fig:sm4ii}
\end{figure}

\begin{figure}[t]
  \begin{center}
    \includegraphics[scale=0.75]{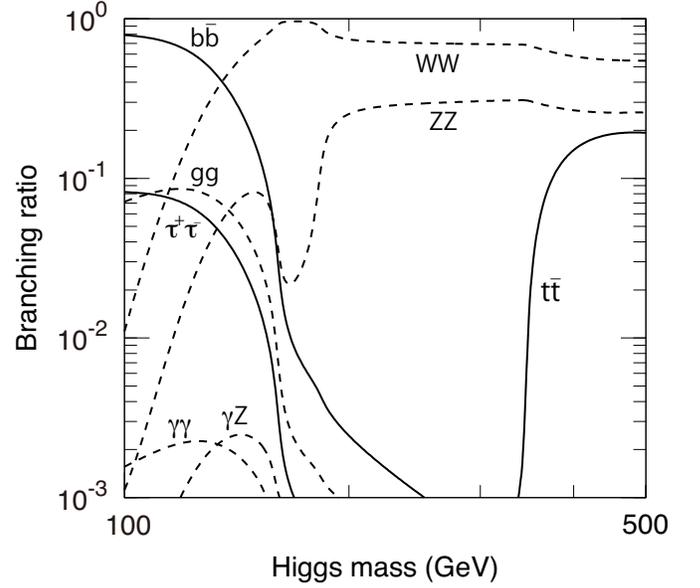}
  \end{center}
  \caption{Branching ratios of the Higgs decay in the standard model.}
  \label{fig:sm}
\end{figure}

In Fig.~\ref{fig:hgamgam}, we plot the ratio of $\Gamma(h\rightarrow
\gamma\gamma)$ to its value in standard model. Here we take
$m_E'=m_E''=100~{\rm GeV}$ and $200~{\rm GeV}$. Since fourth
generation quark contribution typically is negligible, we take the
$M_{Q,U,D}\rightarrow \infty$ limit for simplicity. We find that the
decay width is about $30$-$40$\% of its standard-model value in the
parameter region $115~{\rm GeV} \le m_h \le 150~{\rm GeV}$, in
which $h\rightarrow \gamma \gamma$ is a viable discovery channel for
the Higgs boson.  This is due to the fact that in the loop diagrams
the heavy fermion and $W$ boson contributions interfere
destructively. We also find that the ratio changes by less than about
20\%, depending on the choice of $m_E'$ and $m_E''$ at fixed $m_h$.

The partial decay width of the mode $h\rightarrow \gamma Z$ does not
exhibit this dramatic effect.  We have checked that  the ratio of
the partial decay width of this mode to its standard-model value is
$\simeq 1$.  Finally it is obvious that partial decay width for
$h\rightarrow gg$ does not change at all in the limit
$M_{Q,U,D}\rightarrow \infty$. Therefore, although the Higgs
production rate is almost unchanged from the value predicted in
standard model, the branching ratio for $h\rightarrow \gamma \gamma$
is reduced significantly. This is the outstanding feature
of this model and can be tested at the LHC.

To understand the impact of the fourth generation on Higgs decay
further, we give the Higgs decay branching ratios in
Figs.~\ref{fig:sm4i} and \ref{fig:sm4ii}. Here we take
$m_E'=m_E''=100~{\rm GeV}$ and $m_{N}'=m_{N}''=100~{\rm GeV}$ in
Fig.~\ref{fig:sm4i} and $m_E'=m_E''=100~{\rm GeV}$ and
$m_{N}'=m_{N}''=70~{\rm GeV}$ in Fig.~\ref{fig:sm4ii}.  In the
calculation, we take the limit $M_{Q,U,D}\rightarrow \infty$ as in the
previous plot, and utilize HDECAY package~\cite{Djouadi:1997yw}.  Here
off-shell decays of fourth generation leptons are not considered. For
comparison, we also show the branching ratios in standard model in
Fig.~\ref{fig:sm}.  The decay channels of the standard-model fermion
pairs ($t\bar{t}$, $b{\bar b}$ and $\tau^+ \tau^-$) are shown in solid
lines, those of gauge bosons ($W^+W^-$, $ZZ$, $\gamma\gamma$, $\gamma
Z$ and $gg$) are in dashed lines and those of the fourth generation
leptons pairs ($ e'^{+} e'^{-}$, $e''^{+}e''^{-}$, $\nu' \bar{\nu}'$
and $\nu'' \bar{\nu}''$) are in dot-dashed lines. The line shows the
branching ratio for $e'^+e'^-$ plus $e''^+e''^-$ (${\nu}'\bar{\nu}'$
plus ${\nu}''\bar{\nu}''$) with the sum denoted just by ``$e'^+e'^-$''
(``${\nu}'\bar{\nu}'$'').  We omit $c\bar{c}$ for simplicity of
presentation in those plots. In Fig.~\ref{fig:sm4i}, the branching
ratios are very similar to those in the standard model when $m_h
<200~{\rm GeV}$, except for $h\rightarrow \gamma \gamma$. In the mass
parameter region $m_h >200~{\rm GeV}$, we find that the branching
ratios for $h \rightarrow W^+W^-$ and $ZZ$, which are the main decay
modes, are reduced due to the appearance of the new decay channels,
$h\rightarrow e'^{+} e'^{-}$, $e''^{+}e''^{-}$, $\nu' \bar{\nu}'$ and
$\nu'' \bar{\nu}''$. For example, the branching ratio for $h
\rightarrow W^+W^-$ $(ZZ)$ is 71\% (72\%) of the standard-model value
for $m_h=300~{\rm GeV}$.  In Fig.~\ref{fig:sm4ii}, it is seen that
branching ratios become quite different from those in the standard
model especially around $m_h \sim 150~{\rm GeV}$.  Because of the new
decay channels, the branching ratio for $h\rightarrow W^+W^-$ turns
out to be 27\%, 74\% and 73\% of the standard-model value for
$m_h=150~{\rm GeV}$, $200~{\rm GeV}$ and $300~{\rm GeV}$,
respectively. When $m_h \sim 150~{\rm GeV}$, the Higgs decays mostly
to the fourth generation neutral lepton pairs. Such neutral lepton
pairs are observed as large missing transverse momentum when $\nu'$ or
$\nu''$ does not decay in the detector.  Otherwise, they would be
followed by decay to the standard-model leptons. The decay channels
$e'^{+} e'^{-}$ and $e''^{+}e''^{-}$ are also interesting. These
leptons subsequently decay to the off-shell $W$ boson and $\nu'$ (or
$\nu''$). (The case where the Higgs decays to a stable chiral fourth
generation neutrino pair is previously studied, {\it e.g.},
Ref.~\cite{Belotsky:2002ym}. See also recent
work~\cite{Keung:2011zc}.)

\section{IV. Higgs mass bounds}

As we described in the Introduction, the new generation fermions
affect the running coupling constants in the Yukawa terms and the
Higgs potential.  It is well known that the Yukawa coupling of a new
fermion has a Landau pole not very far above the weak scale when the
additional fermion gets its mass from the Higgs VEV and it is much
heavier than the top quark. Also it is known that such a Yukawa
coupling may cause an instability of the Higgs potential or Landau
pole of the Higgs
self-coupling~\cite{Cabibbo:1979ay,Beg:1983tu,Lindner:1985uk,
  Altarelli:1994rb, Casas:1994qy,Hambye:1996wb}.  In this section we
take the fourth generation quarks to have large masses of order the
cutoff of the theory and we evaluate the running of Yukawa coupling
constants for the fourth generation leptons, the top quark Yukawa
coupling, and the Higgs self-coupling. We use this to discuss the lower
and upper Higgs mass bounds which arise, respectively from avoiding
instability of the vacuum and from the Landau pole in the Higgs
potential.

For simplicity we assume $h'_{E}=h''_{E} \equiv h_E$ and
$h'_{N}=h''_{N}\equiv h_N$.  Then the renormalization group equations
(RGEs) for the lepton Yukawas and the top Yukawa are
\begin{eqnarray}
16\pi^2 \mu \frac{\partial h_E}{\partial \mu}
&=&- h_E\left(\frac{9}{4}g_2^2+\frac{15}{4}g_1^2\right)  
+ \frac{7}{2}h^3_E
\nonumber \\
&&+  h_E\left( 3 y_t^2 + \frac{1}{2} h^2_N \right),
\\
16\pi^2 \mu \frac{\partial h_N}{\partial \mu}
&=&- h_N\left(\frac{9}{4}g_2^2+\frac{3}{4}g_1^2\right)  
+ \frac{7}{2}h^3_N
\nonumber \\
&&+  h_N\left( 3 y_t^2 + \frac{1}{2} h^2_E \right),
\\
16\pi^2 \mu \frac{\partial y_t}{\partial \mu}
&=&-y_t\left(8g_3^2+\frac{9}{4}g_2^2+\frac{17}{12}g_1^2\right)  
+ \frac{9}{2}y_t^3
\nonumber \\
&&+  y_t\left(2 h^2_E+2 h^2_N \right).
\end{eqnarray}
(For formulae we refer to Ref.~\cite{Hashimoto:2010at}. See also
Refs.~\cite{Cheng:1973nv,Machacek:1981ic}).  Here $g_3$, $g_2$ and
$g_1$ are gauge coupling constants of $SU(3)_c$, $SU(2)$ and $U(1)_Y$,
respectively, and $\mu$ is the renormalization scale.  In our
evaluation we neglect all other quark and lepton Yukawa
couplings. RGEs of the gauge couplings are,
\begin{eqnarray}
16 \pi^2 \mu \frac{\partial g_i}{\partial \mu}
= -b_i g_i^3,
\end{eqnarray}
with
\begin{eqnarray}
b_1 &=& -\frac{2}{3}\left(\frac{3}{2}n_L+\frac{11}{6}n_Q\right) 
-\frac{1}{6}n_H,
\\
b_2 &=& \frac{22}{3} -\left(\frac{1}{3}n_L+n_Q\right)
-\frac{1}{6}n_H,
\\
b_3 &=& 11-\frac{4}{3}n_Q.
\end{eqnarray}
Here $n_L$ and $n_Q$ are the number of generations of leptons and
quarks, and $n_H$ is the number of the Higgs doublets.  In our model
$n_H=1$, $n_L=5$ (three generation plus fourth generation and its
mirror) and $n_Q=3$, assuming that heavy fourth generation quarks have
masses of order the cutoff of the theory so they do not contribute to
the running of the gauge couplings. For the Higgs sector we write
Higgs potential as
\begin{eqnarray}
V^H = -\mu_H^2 |H|^2+\lambda |H|^4,
\end{eqnarray}
so that the Higgs mass is, 
\begin{equation}
\label{hmass}
m_h=\sqrt{2 \lambda} v. 
\end{equation}
In this convention, RGE for $\lambda$ is given by
\begin{eqnarray}
&&16\pi^2 \mu \frac{\partial \lambda}{\partial \mu}
= 24 \lambda^2 -3 \lambda(3g_2^2+g_1^2)
\nonumber \\
&& + 4\lambda\left[3y_t^2 +2(h^2_E+h^2_N)\right]
-2 \left[ 3y_t^4 +2(h^4_E+h^4_N) \right]
\nonumber \\
&&+\frac{3}{8} \left[2g_2^4+(g_2^2+g_1^2)^2\right].
\end{eqnarray}

\begin{figure}[t]
  \begin{center}
    \includegraphics[scale=1.4]{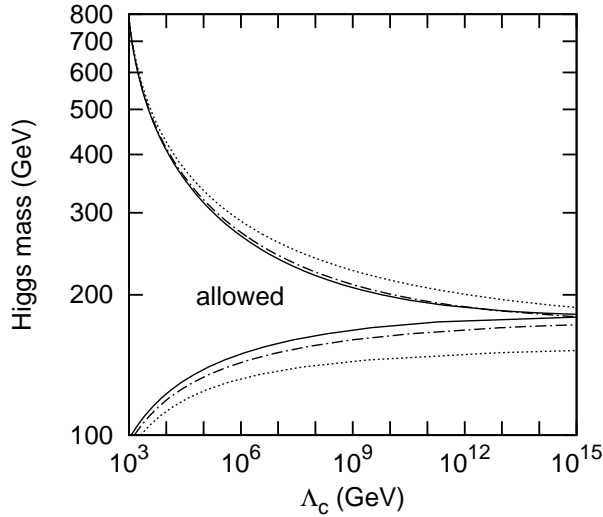}
  \end{center}
  \caption{Upper and lower bounds on the Higgs mass as a function of
    cutoff. The results are shown in solid (dot-dashed) lines for the
    case of $m'_{E}=m''_{E}=100~{\rm GeV}$ and $m'_{N}=m''_{N}=100$
    ($70)~{\rm GeV}$.  Dotted lines show the results in the standard
    model.  }
  \label{fig:HiggsLimMe100}
\end{figure}

\begin{figure}[t]
  \begin{center}
    \includegraphics[scale=1.4]{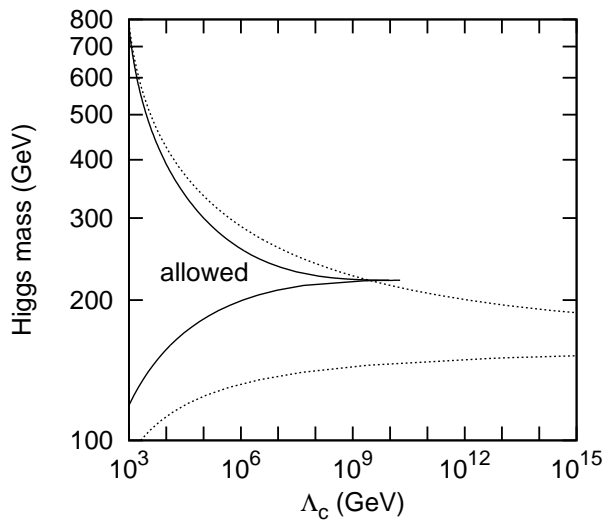}
  \end{center}
  \caption{The same plot as Fig.~\ref{fig:HiggsLimMe100} except for
    taking $m'_{E}=m''_{E}=m'_{N}=m''_{N}=150~{\rm GeV}.$ }
  \label{fig:HiggsLimMe150}
\end{figure}

\begin{figure}[t]
  \begin{center}
    \includegraphics[scale=1.4]{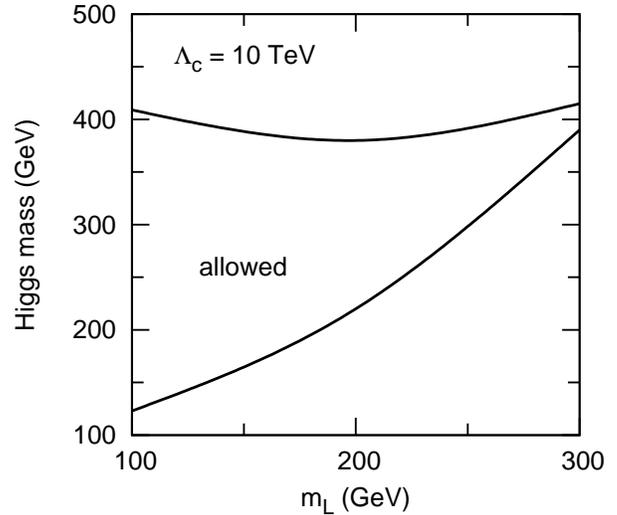}
  \end{center}
  \caption{Allowed region on the Higgs mass vs.~fourth generation
    masses for $\Lambda_c=10~{\rm TeV}$. Here we take
    $m'_{E}=m''_{E}=m'_{N}=m''_{N} \equiv m_L$. }
  \label{fig:MLMh}
\end{figure}

Solving those RGEs, we derive the Higgs mass bounds by imposing
$0<\lambda(\mu)<2 \pi$ \footnote{The upper range is not the location
  of the Landau pole but rather the end of the weak coupling
  region. In practice provided one uses one loop running, this
  distinction is not important numerically.}, with the Higgs mass
determined by Eq.~(\ref{hmass}) using the coupling $\lambda$ evaluated
at the Higgs mass.  At some scale the condition $0<\lambda(\mu)<2 \pi$
can not be satisfied and we interpret this scale as a cutoff for the
model, $\Lambda_c$. Thus, the Higgs mass bounds are given as a
function of the cutoff. The condition $\lambda(\mu)>0$ gives the lower
bound for the Higgs mass, while the condition $\lambda(\mu) <2\pi$
gives the upper bound.  Numerical results are given in
Figs.~\ref{fig:HiggsLimMe100} and \ref{fig:HiggsLimMe150}. Here we
take $m'_{E}=m''_{E}=m'_{N}=m''_{N}=100~{\rm GeV}$ ($150~{\rm GeV}$)
in Fig.~\ref{fig:HiggsLimMe100} (Fig.~\ref{fig:HiggsLimMe150}). In
Fig.~\ref{fig:HiggsLimMe100} we also plot the result when
$m'_{E}=m''_{E}=100~{\rm GeV}$ and $m'_{N}=m''_{N}=70~{\rm GeV}$ are
chosen using a dot-dashed line.  In the plots, we also give the result
in standard model using a dotted line.  In the first case ({\it i.e.},
Fig.~\ref{fig:HiggsLimMe100}) we have checked that the Yukawa
couplings do not have Landau poles up to the Planck scale, while in
the second case ({\it i.e.}, Fig.~\ref{fig:HiggsLimMe150}) the top
quark Yukawa has a Landau pole around $\mu \sim 10^{10}~{\rm GeV}$. In
Fig.~\ref{fig:HiggsLimMe100}, $m_h\sim 180$ $(170{\text -}180~{\rm
  GeV})$ is indicated for $m'_{N}=m''_{N}=100~(70)~{\rm GeV}$ when the
cutoff of the theory is about $10^{15}~{\rm GeV}$. When the Higgs is
lighter, the cutoff is significantly reduced. From the numerical
calculations, we find $\Lambda_c \simeq 4.3~(6.2)~{\rm TeV}$,
$26~(54)~{\rm TeV}$ and $1.2~(8.6) \times 10^{3}~{\rm TeV}$ for lower
bounds, $m_h=115~{\rm GeV}$, $130~{\rm GeV}$ and $150~{\rm GeV}$ when
$m'_{N}=m''_{N}=100$ ($70)~{\rm GeV}$. In
Fig.~\ref{fig:HiggsLimMe150}, $m_h\sim 210~{\rm GeV}$ is implied when
cutoff of the theory is near the Landau pole of the Yukawa couplings,
{\it i.e.}, $\sim 10^{10}~{\rm GeV}$. Similarly to the previous
result, we obtained $\Lambda_c \simeq 8.3 \times 10^2~{\rm GeV}$,
$1.8~{\rm TeV}$ and $8.2~{\rm TeV}$ for lower bounds $m_h=115~{\rm
  GeV}$, $130~{\rm GeV}$ and $150~{\rm GeV}$.  In Fig.~\ref{fig:MLMh},
the allowed region for the Higgs mass vs.~fourth generation lepton
masses is given for a fixed cutoff of $10~{\rm TeV}$. Here we take
$m'_{E}=m''_{E}=m'_{N}=m''_{N}\equiv m_L$.  We found $120~{\rm
  GeV}\lesssim m_h\lesssim 400~{\rm GeV}$ and $m_L\lesssim m_h$ is the
allowed region. This is consistent with what is expected from previous
works where a similar analysis was performed for chiral fourth
generation scenario~\cite{Hashimoto:2010at}.

Finally we note that evaluation of the Higgs mass bound has
theoretical uncertainty coming, for example, from matching conditions
of fermion and the Higgs sector at the low energy
boundary~\cite{Hambye:1996wb}. The allowed Higgs mass region may
change due to this; however it is shown in Ref.~\cite{Hambye:1996wb}
that this uncertainty is less than $\sim 10~{\rm GeV}$ in the standard
model. We do not estimate this kind of uncertainty, expecting a
similar order of uncertainty in our case.

\section{V. Concluding Remarks} 

Although we observe three chiral generations of quarks and leptons,
there is no established physical principal that fixes the number of
generations.  In this paper we consider an additional vector-like
generation. Within this framework, we focus on a scenario where fourth
generation quarks gets large masses without the Higgs VEV, while
fourth generation lepton masses are determined by weak symmetry
breaking. Then quark sector Yukawa couplings do not develop Landau
poles near the weak scale. We have studied Higgs properties in this
scenario. We found that the new leptons reduce the branching ratio for
$h \rightarrow \gamma \gamma$ to about $30\%$ of its standard-model
value. Furthermore the Higgs production rate at the LHC is very near
its standard-model value if the new fourth generation quarks are much
heavier than the weak scale. We have also examined the upper and lower
limits on the Higgs mass in this model from the condition that all the
Yukawa coupling constants and the Higgs self-coupling are free of
Landau poles and that the familiar weak symmetry breaking vacuum is
stable.  We found when cutoff of the theory is about $10^{15}~{\rm
  GeV}$ then $m_h\sim 175~{\rm GeV}$ and fourth generation lepton
masses should not be greater than about $100~{\rm GeV}$.  When cutoff
is around $10~{\rm TeV}$, $120~{\rm GeV}\lesssim m_h\lesssim 400~{\rm
  GeV}$ with fourth generation lepton masses being roughly less than
$m_h$ in the allowed region.

\subsection*{Acknowledgment}
The work was supported in part by the U.S. Department of Energy under
Contract No. DE-FG02-92ER40701, and by the Gordon and Betty Moore
Foundation.


\end{document}